\documentclass[amsmath,amssymb,aps,twocolumn,pre]{revtex4-2}

\usepackage{graphicx}
\usepackage{dcolumn}
\usepackage{bm}
\usepackage{hyperref}
\usepackage{multirow}
\usepackage[table]{xcolor}
\usepackage{ulem}

\usepackage{xcolor}
\newcommand{\ea}[1]{\textcolor{black}{#1}}      % for Emilien
\begin{document}

\preprint{APS/123-QED}

\title{Origin of geometric cohesion in non-convex granular materials: interplay between interdigitation and rotational constraints enhancing frictional stability.}
\author{Jonathan Barés \textsuperscript{1}}%
\email{jonathan.bares@umontpellier.fr}

\author{Arnaud Regazzi \textsuperscript{3}}%
\email{arnaud.regazzi@mines-ales.fr}

\author{David Aponte$^{1,2}$} 
\email{david.aponte@umontpellier.fr}

\author{Sylvain Buonomo \textsuperscript{3}}%
\email{sylvain.buonomo@mines-ales.fr}

\author{Mathieu Renouf \textsuperscript{1}}%
\email{mathieu.renouf@umontpellier.fr}

\author{Nicolas Estrada \textsuperscript{2}}
\email{n.estrada22@uniandes.edu.co}

\author{Emilien Azéma \textsuperscript{1,4}}%
\email{emilien.azema@umontpellier.fr}
 
\affiliation{$^{1}$ LMGC, Université de Montpellier, CNRS, Montpellier, France}
\affiliation{$^{2}$ Departamento de Ingeniería Civil y Ambiental, Facultad de Ingeniería, Universidad de los Andes, Bogotá, Colombia}
\affiliation{$^{3}$ LMGC, Univ Montpellier, IMT Mines Ales, CNRS, Ales, France}
\affiliation{$^{4}$ Department of Civil, Geological, and Mining Engineering, Polytechnique Montréal, Montréal, Canada}

\date{\today}
\begin{abstract}
We present a series of experiments investigating the local microstructure of cylindrical piles composed of highly concave particles. By systematically varying particle geometry -- from spheres to strongly non-convex polypods -- as well as frictional properties and the number of branches, we explore how these parameters, together with the preparation protocol, shape the internal structure of the system. Using X-ray tomography combined with a dedicated image-analysis pipeline, we accurately extract the position, orientation, and contacts of every particle in each pile. This allows us to quantify the evolution of key structural observables as a function of particle geometry and preparation method. In particular, we measure the distributions of local packing fraction, coordination number, number of neighbors, and contact locations, along with particle-particle positional and orientational correlations. More importantly, we construct a new stability indicator that correlates perfectly with the observed pile stabilities, enabling us to identify the fundamental mechanisms responsible for \textit{geometrically induced cohesion} in granular systems composed of non-interlocking particle shapes: interdigitation, rotational constraint, friction-mediated cohesion, and the ability of a pile to re-stabilize.

\end{abstract}


\maketitle

\tableofcontents

\section{Introduction}
Granular matter is widely used in technical applications because it can be easily manipulated and transported in its fluid-like phase, while also being exploited for its strength in its solid-like phase. The jamming transition -- the change between these two phases -- as well as the stability of the jammed state have therefore been extensively studied over the past decades across many domains. Recently, a new family of granular systems composed of non-convex grains, also called \textit{metagrains}, has emerged \cite{dierichs2015_ad,dierichs2016_gm,cantor2022_pp}. 
These systems make it possible to form extremely stable jammed structures while remaining easy to manipulate in their fluid phase. The seminal work of S.~V.~Franklin \cite{franklin2012_pt} first demonstrated the existence of \textit{geometrically induced cohesion} in metagranular structures. This means that in some specific cases mechanical stability of a granular system arises purely from the shape and arrangement of its particles, without any attractive forces.
This concept has recently been extended to \textit{geometrically induced consistency} \cite{wang2024_prr,wang2025_pre}, describing structures that, while consistent, are still capable of flowing. Such cohesion has been reported in a wide variety of engineered systems with diverse particle shapes, but it is also observed in natural systems, with an equally broad zoology of particle geometries, ranging from snowflakes \cite{libbrecht2005_rpp} to sponge spicules \cite{lukowiak2022_jm,sethmann2008_mic}.

Among the most extensively studied shapes are U-, C-shaped, and staple-like grains, which have been shown, both experimentally and numerically, to form structures resistant to tensile forces \cite{franklin2014_epl,pezeshki2025_jmps,sohn2025_gm,kim2025_axv,murphy2016_gm,marschall2015_gm}. In these cases, cohesion arises from particle \textit{interlocking}/\textit{entanglement}, where grains literally form chains -- meaning that two particles can be pulled apart yet still remain hooked to one another.
However, cohesion is also observed in systems composed of particles that cannot interlock or entangle. A prominent example is given by polypod particles, which can form remarkably stable piles without any external cohesive forces. These particles consist of elongated, straight, rigid branches radiating from a central core. They were popularized by the seminal work of K.~Dierichs and A.~Menges, who demonstrated the potential of metagrains for constructing human-scale architectural structures \cite{dierichs2015_ad,dierichs2016_gm,keller2016_gm}. More quantitative studies have since investigated these systems, characterizing pile stability \cite{zhao2016_gm,aponte2025_gm}, their resistance to external loading \cite{zhao2016_gm}, vibrations \cite{zhao2017_epj,luo2025_arx}, and shear deformation \cite{zhao2020_pre,li2025_part}.

Although the stability of such structures is now well documented, the origin of geometrically induced cohesion remains elusive. 
It arises from the competition between two phenomena: (\textit{i}) particle \textit{interdigitation} which determines how closely two particles can approach each other given the spatial extent of their branches, and (\textit{ii}) inter-particle \textit{rotational constraint} which limits the relative rotation of neighboring particles due to branch-branch interactions.
Two-dimensional analyses have been carried out to probe, at the grain scale, the mechanisms responsible for stability. Both experimental measurements \cite{zheng2017_epj} and numerical simulations \cite{aponte2024_pre} in 2D have shown that system connectivity -- particularly the ability to form multi-contact interactions that hinder the relative rotation of grains -- is the main factor underlying pile stability. However, as emphasized by N.~A.~Conzelmann \textit{et al.} \cite{conzelmann2020_pre} in the context of concave granular matter, 2D results cannot be directly extrapolated to 3D systems, as proven for the case of force chains. Extending such local-scale analyses to 3D is extremely challenging, as visualization and quantitative tracking are far more complex. In the case of granular chains forming 3D packings, E.~Brown \textit{et al.} \cite{brown2012_prl} used X-ray tomography, and preliminary attempts have been made for hexapod systems \cite{bares2024_git,luo2025_arx}.

In this study, we take a step further in exploring the origin of \textit{geometrically induced cohesion} in polypod systems. By varying particle friction (from nearly frictionless to highly frictional), particle geometry (from spherical to highly concave polypods, see Fig.~\ref{fig_expe}a), the number of branches, and preparation protocols, we investigate the onset of stability and its relation to key observables in granular packings.
%NE: I have the impression that we are presenting fig. b first and fig. a later. This is not intuitive, and suggests the reader that he/she skipped fig. a without noticing. These figures are really pretty. Maybe, we could separate fig. b from figs. a and c ...
%NE: Another detail: We introduce fig. b, but when one goes to read the caption, one finds accronyms such as HDPE and PA12, which are not yet defined.
%Ne: Maybe, we should avoid presenting the figure here, since it is well presented and explained in next session.
% Agree: does it is necessary to invok the figure here?
The work is conducted on model pile systems composed of polypods -- model particles known for their ability to form extremely stable structures. Our findings reveal that 
%\ar{I'm not familiar with the habits in the field but I find it very unusual to present results in the introduction} \jb{I do not like it neither but it is what is done. It is supposed to be a teaser to insite you to read the rest} 
the extreme behavior of these materials manifests at the local scale in the form of very low-density regions and unusually high packing fractions. We further show that stability is not linked to any specific long-range ordering or crystallinity \cite{stannarius2022_gm,meng2020_part}, but rather tends to emerge in fully disordered structures. Crucially, pile stability appears to be governed by the system's capacity to remain \textit{reconformable}: that is, to form a large number of potentially reconfigurable contact chains. This reconformability enhances the ability of the system to rebalance and remain stable after small perturbations.

This article is structured as follows. In Section~\ref{sec_expe}, we describe the experimental setup and model particles. Section~\ref{sec_image} details the X-ray data post-processing procedure and the extraction of local observables for each particle. These observables are then analyzed in Section~\ref{sec_result}, where we explore their correlation with pile stability. Finally, Section~\ref{sec_ccl} discusses the implications of these results and offers concluding remarks.

\section{Experimental setup and material} \label{sec_expe}

\begin{figure}[b!]
    \centering
    \includegraphics[width=0.99\columnwidth]{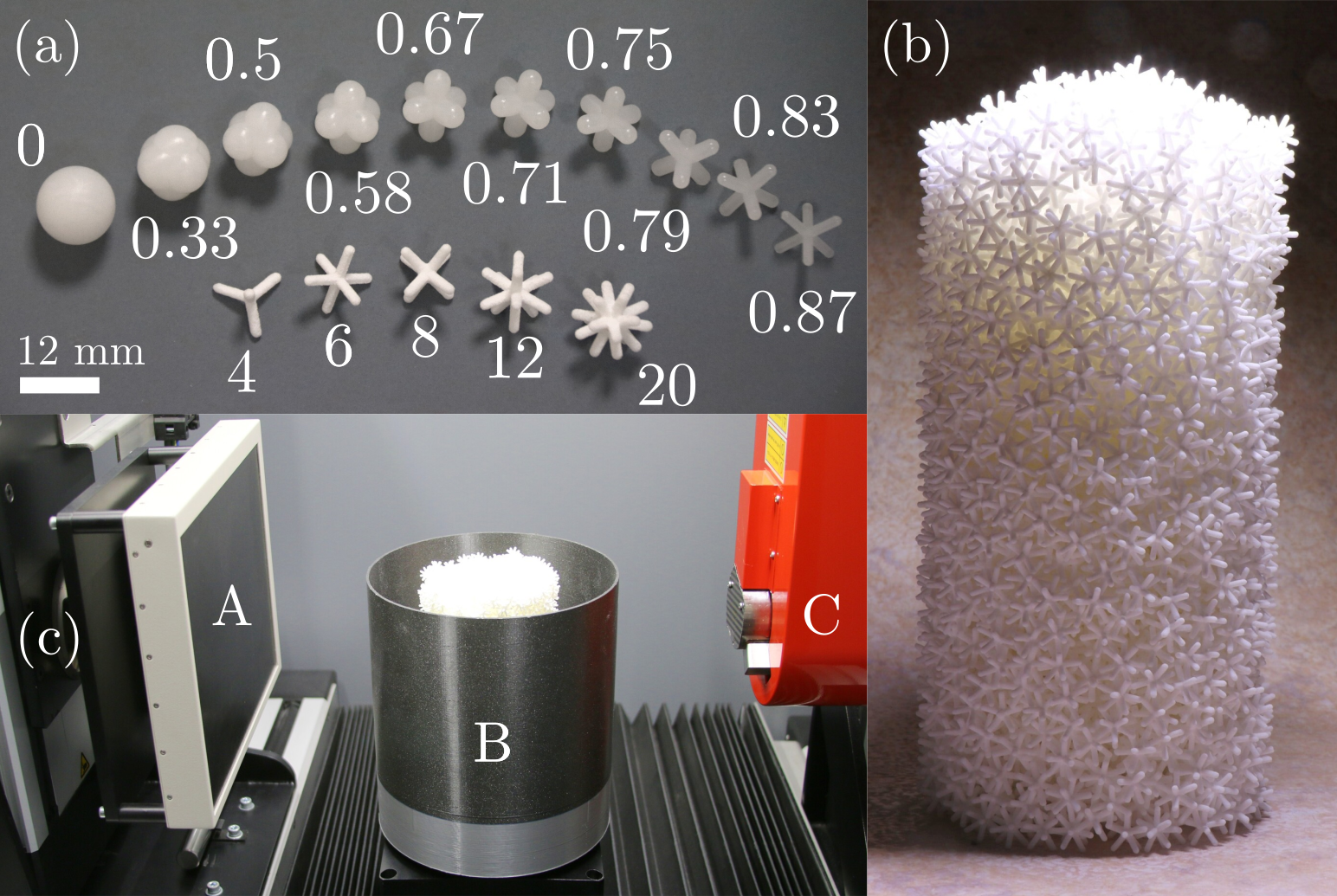}
    \caption{Experiment. (a) Top: High-density polyethylene (HDPE) particles with $4$ branches and concavity $\eta$, ranging from $0$ (sphere) to $0.87$ (hexapod with branch thickness of $1.5$ mm). Concavity values are given on the picture. Bottom:  Polyamide 12 (PA12)  particles with a fixed branch thickness of $1.5$ mm and a varying number of branches from $4$ to $20$. The number of branches is given on the picture for each particle. (b) Cylindrical pile with a diameter of $15$ cm composed of PA12 particles with $12$ branches. (c) PA12 sample with $20$ branches positioned in the X-ray microtomograph. A: X-ray CCD detector collecting the images; B: pile sample contained in a plastic bucket placed on the rotation stage; C: X-ray tube. }
    \label{fig_expe}
\end{figure}

The experimental setup involves forming cylindrical piles of monodisperse particles through various methods and scanning them with a X-ray microtomograph to characterize their microstructure. The cylindrical piles (see, for example, Fig.~\ref{fig_expe}b) are composed of monodisperse particles of different shapes and materials, as illustrated in Fig.~\ref{fig_expe}a. All particles are inscribed within a sphere of diameter $d = 12$ mm. In the top part of Fig.~\ref{fig_expe}a, the particles consist of $6$ spherocylindrical branches extending toward the center of the faces of a regular cube. The branch radius, denoted $r_0$, decreases gradually from $6$ mm (sphere) to $0.75$ mm. These particles are fabricated from High-Density Polyethylene (HDPE) and Ethylene Propylene Diene Monomer (EPDM) using injection molding with a custom-made mold \cite{wang2024_prr,wang2025_pre,aponte2025_gm}, producing clusters of particles. By contrast, the particles shown in the bottom part of Fig.~\ref{fig_expe}a are manufactured using Selective Laser Sintering (SLS) with Polyamide 12 (PA12). In this case, the particle diameter $d$ and branch radius $r_0$ are both fixed ($d = 12$ mm and $r_0 = 1.5$ mm), while the number of branches, $n_b$, varies from $3$ to $20$. The configurations include branch arrangements derived from the five Platonic solids ($n_b \in [4, 6, 8, 12, 20]$). The selected materials and loading ensure that the particles behave as rigid, non-deformable bodies. Their mechanical properties are summarized in Tab.~\ref{tab_mat}.

\begin{table}[h]
    \centering
    \begin{tabular}{|l|c|c|c|}
        \hline
        \multirow{2}{*}{Material} & Young's     & Coulomb friction  & Density \\
                                  & modulus $E$ & coefficient $\mu$ & $\rho$  \\
        \hline
        EPDM & 80 MPa & 0.5 & 0.91 kg·m$^{-3}$ \\ \hline
        HDPE & 1300 MPa & 0.2 & 0.96 kg·m$^{-3}$ \\ \hline
        PA12 & 1650 MPa & 0.8 & 0.93 kg·m$^{-3}$ \\
        \hline
    \end{tabular}
    \caption{Main properties of the materials used for the particles}
    \label{tab_mat}
\end{table}

To quantify the particle geometry, a concavity parameter, denoted by $\eta$, is defined as $\eta = 1 - 2r_0/d$ . For HDPE and EPDM particles, this parameter varies within the range $\eta \in {0, 0.33, 0.5, 0.58, 0.67, 0.75, 0.79, 0.83, 0.87}$. In contrast, for PA12 particles (see Fig.~\ref{fig_expe}a), $\eta$ is fixed at $0.87$. This geometrical parameter is adapted from the definition introduced in \cite{saintcyr2012_epl}.

Three types of samples are prepared: (\textit{i}) free-standing cylindrical piles, (\textit{ii}) confined columns without vibration, and (\textit{iii}) confined columns with vibration. In the latter two cases, the particles are contained within a rigid cylindrical mold. Regardless of the sample type, the initial preparation procedure is the same: particles are gradually poured into a cylinder with a diameter of $10$ cm until reaching a height of approximately $15$ cm. For \textit{confined non-vibrated} samples, the process ends at this stage. For \textit{confined vibrated} samples, the column is subsequently subjected to horizontal vibration for approximately $30$ seconds using an eccentric device similar to the one described in \cite{zhao2016_gm}. As recently observed by Luo \textit{et al.} \cite{luo2025_arx}, $30$ seconds is enough to reach a steady state. For \textit{free-standing} samples, the cylindrical mold is carefully removed by gently pulling apart the two halves of the split cylinder. The resulting self-supporting structure is then placed in an X-ray microtomograph, as illustrated in Fig.~\ref{fig_expe}c. To prevent particle loss during scanning, the sample is placed inside a plastic bucket positioned on the rotary stage of the device. The bucket has negligible attenuation of the X-ray beam and does not interfere with imaging.

X-ray scans are performed using an EasyTom 150 system from RX-Solutions (Chavanod, France) equipped with a PaxScan$^{\mbox{\scriptsize{\textregistered}}}$ 2520DX $1920 \times 1536$~px$^2$ CCD detector from Varex Imaging (Salt Lake City, UT, USA). For all scans, the X-ray source is set to $50$ kV and $500$ µA. The distance between the sample and the CCD detector is set to $391$ mm, while the distance between the sample and the X-ray source is set to $195$ mm, resulting in a magnification factor of $2.01$ and a voxel size of $63.3$ µm. A carbon plate is placed between the sample and the source to homogenize the X-ray beam. Each scan comprises 1440 projections, with each projection averaged over five acquisitions to enhance the signal-to-noise ratio. The reconstructed slices are generated from these projection images using the X-Act software provided by RX-Solutions. Each sample is scanned twice with a vertical offset to cover the entire column height, and the two datasets are subsequently stitched together using the same software. The complete scanning and reconstruction process takes approximately one hour per sample. A total of $20$ experiments are conducted, with their specific characteristics listed in Tab.~\ref{tab_list}.
These experiments allow for the investigation of the effects of interparticle friction, particle concavity and number of branches, packing vibration, and confinement on the microstructure of cylindrical piles. Two of the experiments were repeated to verify the reproducibility of the statistical observables within their associated error bars.

Alongside the scanning experiments, the stability of the columns for each preparation protocol and particle type was also tested. We used the same procedure as in \cite{zhao2016_gm,aponte2025_gm}. Columns were formed using the same preparation method as for the scanned samples, with or without vibration. After preparation, the confining cylinder was gently removed, causing the column to collapse to varying degrees. All particles that ended up outside the original cylinder footprint were collected and weighed, and their mass was compared to the total particle mass. The ratio of fallen mass to total mass defines the collapse ratio, $\mathcal{R}$, which quantifies the column stability. For each configuration, this measurement was repeated at least five times. The corresponding values are reported in the last column of Tab.~\ref{tab_list}.

\begin{table}
    \centering
    \begin{tabular}{|c|c|c|c|c|c|c|c|}
        \hline
        \rotatebox{70}{Material} & \rotatebox{70}{Friction, $\mu$} & \rotatebox{70}{Branch number, $n_b$} & \rotatebox{70}{Branch diameter, $2r_0$} & \rotatebox{70}{Concavity, $\eta$} & \rotatebox{70}{Confined} & \rotatebox{70}{Vibrated} & \rotatebox{70}{Collapse ratio, $\mathcal{R}$} \\
        \hline
        \multirow{2}{*}{HDPE}  & \multirow{2}{*}{0.2}  & \multirow{2}{*}{6}  & 1.5                  & 0.87                  & yes & no  & 0.71$\pm$0.06                    \\ \cline{4-8}
                               &                       &                     & 1.5                  & 0.87                  & yes & yes & 0.65$\pm$0.04                    \\ \hline
        \multirow{12}{*}{EPDM} & \multirow{12}{*}{0.5} & \multirow{12}{*}{6} & 1.5                  & 0.87                  & yes & no  & 0.52$\pm$0.05                    \\ \cline{4-8}
                               &                       &                     & 1.5                  & 0.87                  & yes & yes & 0.46$\pm$0.03                    \\ \cline{4-8}
                               &                       &                     & 2                    & 0.83                  & yes & no  & \multirow{2}{*}{0.63$\pm$0.07}   \\ \cline{4-7}
                               &                       &                     & 2                    & 0.83                  & yes & no  &                                  \\ \cline{4-8}
                               &                       &                     & 2.5                  & 0.79                  & yes & no  & \multirow{2}{*}{0.68$\pm$0.06}   \\ \cline{4-7}
                               &                       &                     & 2.5                  & 0.79                  & yes & no  &                                  \\ \cline{4-8}
                               &                       &                     & 3                    & 0.75                  & yes & no  & 0.71$\pm$0.04                    \\ \cline{4-8}
                               &                       &                     & 4                    & 0.67                  & yes & no  & 0.75$\pm$0.04                    \\ \cline{4-8}
                               &                       &                     & 5                    & 0.58                  & yes & no  & 0.77$\pm$0.03                    \\ \cline{4-8}
                               &                       &                     & 6                    & 0.5                   & yes & no  & 0.79$\pm$0.02                    \\ \cline{4-8}
                               &                       &                     & 8                    & 0.33                  & yes & no  & 0.82$\pm$0.02                    \\ \cline{4-8}
                               &                       &                     & 12                   & 0                     & yes & no  & 0.89$\pm$0.02                    \\ \hline
        \multirow{6}{*}{PA12}  & \multirow{6}{*}{0.8}  & 4                   & \multirow{6}{*}{1.5} & \multirow{6}{*}{0.87} & no  & yes & 0.33$\pm$0.07                    \\ \cline{3-3} \cline{6-8}
                               &                       & 6                   &                      &                       & yes & no  & 0.27$\pm$0.04                    \\ \cline{3-3} \cline{6-8}
                               &                       & 6                   &                      &                       & no  & yes & 0.22$\pm$0.03                    \\ \cline{3-3} \cline{6-8}
                               &                       & 8                   &                      &                       & no  & yes & 0.12$\pm$0.02                    \\ \cline{3-3} \cline{6-8}
                               &                       & 12                  &                      &                       & no  & yes & 0.07$\pm$0.01                    \\ \cline{3-3} \cline{6-8}
                               &                       & 20                  &                      &                       & no  & yes & 0.06$\pm$0.01                    \\ \hline
    \end{tabular}
    \caption{This table presents the list of experiments performed along with their corresponding input parameters. Two pairs of experiments (EPDM with $\eta = 0.83$ and $\eta = 0.79$) are intentionally duplicated to assess the reproducibility of the results. The last column reports the stability score, $r$, as initially defined in \cite{zhao2016_gm}.}
\label{tab_list}
\end{table}

\section{Image analysis} \label{sec_image}
\begin{figure}[b!]
\centering
\includegraphics[width=0.99\columnwidth]{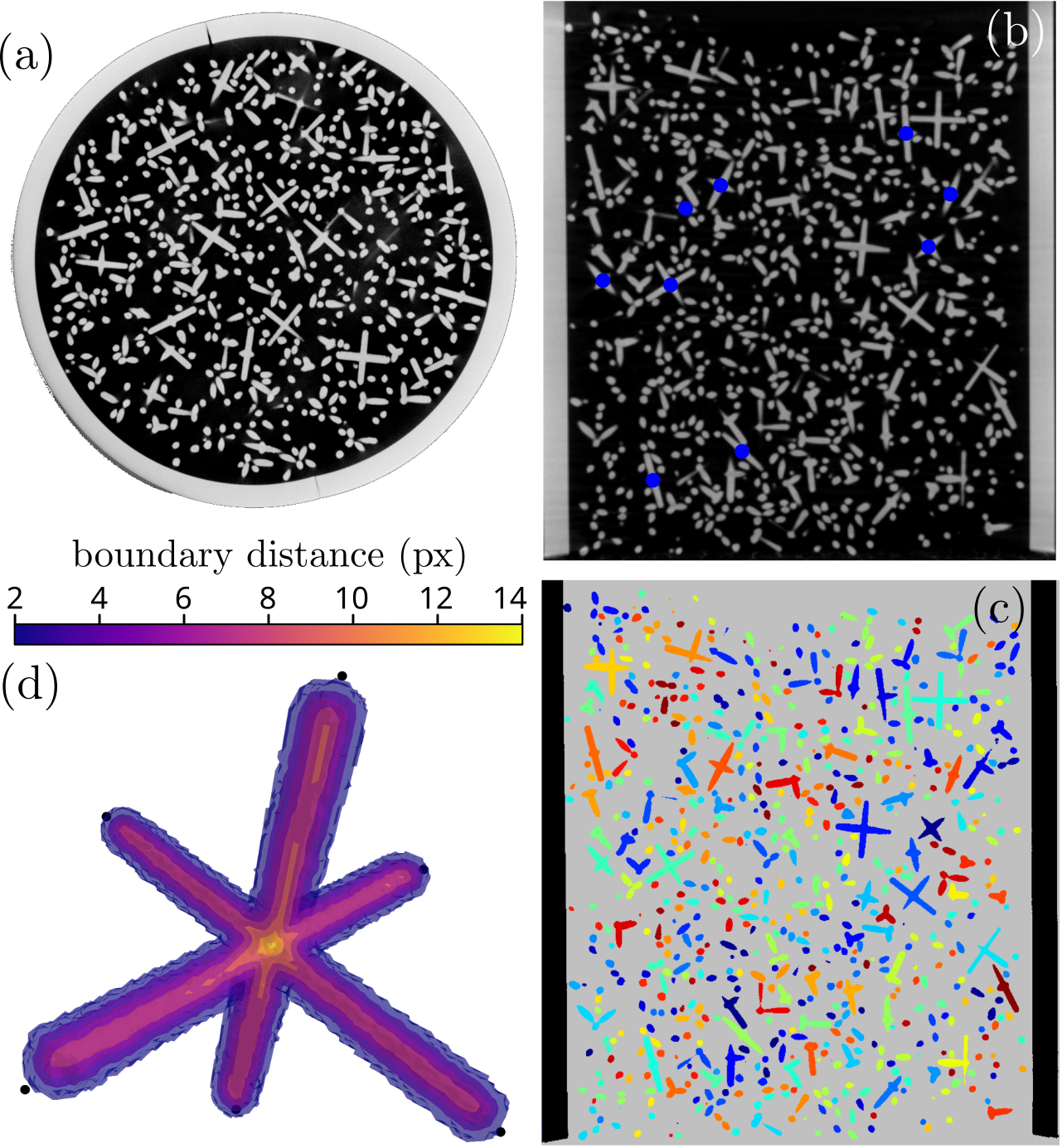}
\caption{Image post-processing. (a) Reconstructed horizontal slices of the X-ray density matrix of a hexapod pile. (b) Vertical slice of the same system, with blue points indicating the roughly measured positions of the particle centers. (c) Segmented version of the same vertical slice, where individual particles are displayed in different colors. (d) 3D iso-value rendering of the boundary-distance matrix of a single particle, with black points marking the measured extremities of its branches. All images are extracted or computed from the X-ray scan of a cylindrical pile composed of EPDM hexapods with a branch cross-section of $1.5$ mm. }
\label{fig_p-pro}
\end{figure}

For each experiment, we obtain a stack of images corresponding to horizontal slices of the system density (see Fig.~\ref{fig_p-pro}a). Using a homemade Python code \cite{bares2024_git} derived from the one already presented by Barés \textit{et al.} \cite{bares2017_epj}, these images are assembled into a 3D matrix of greyscale voxels (8-bit unsigned integer). This matrix is then thresholded to separate the particles (bright regions) from the background (dark regions), yielding a binary 3D representation. By applying successive erosion processes, the particle branches are progressively cropped, leaving small voxel clusters around the particle centers. The barycenters of these clusters provide approximate grain positions (see Fig.~\ref{fig_p-pro}b) and the number of particles.

From these approximate positions, a watershed algorithm is applied to the binarized 3D matrix in order to extract the particles individually (see Fig.~\ref{fig_p-pro}c). Each extracted particle is then processed separately by isolating a sub-3D matrix containing only that particle binarized pattern. On each sub-matrix, an Euclidean distance transform is applied, assigning to every voxel in the particle the distance to the closest particle boundary (see Fig.~\ref{fig_p-pro}d). This produces a new sub-3D matrix in which values are maximal along the backbone of the particle. From this transform, values are interpolated along rays leaving from the approximate particle center in all directions. By locating the maxima of the integrals along these rays -- aligned with the particle branches -- we obtain both the number of branches and their rough orientations. To refine these measurements, we maximize a function defined as the integral of the interpolated matrix values along rays starting from the particle center and extending toward the approximate branch extremities. This procedure yields accurate measurements of both the particle positions and their orientations (see Fig.~\ref{fig_p-pro}d).

\begin{figure}[b!]
\centering
\includegraphics[width=0.99\columnwidth]{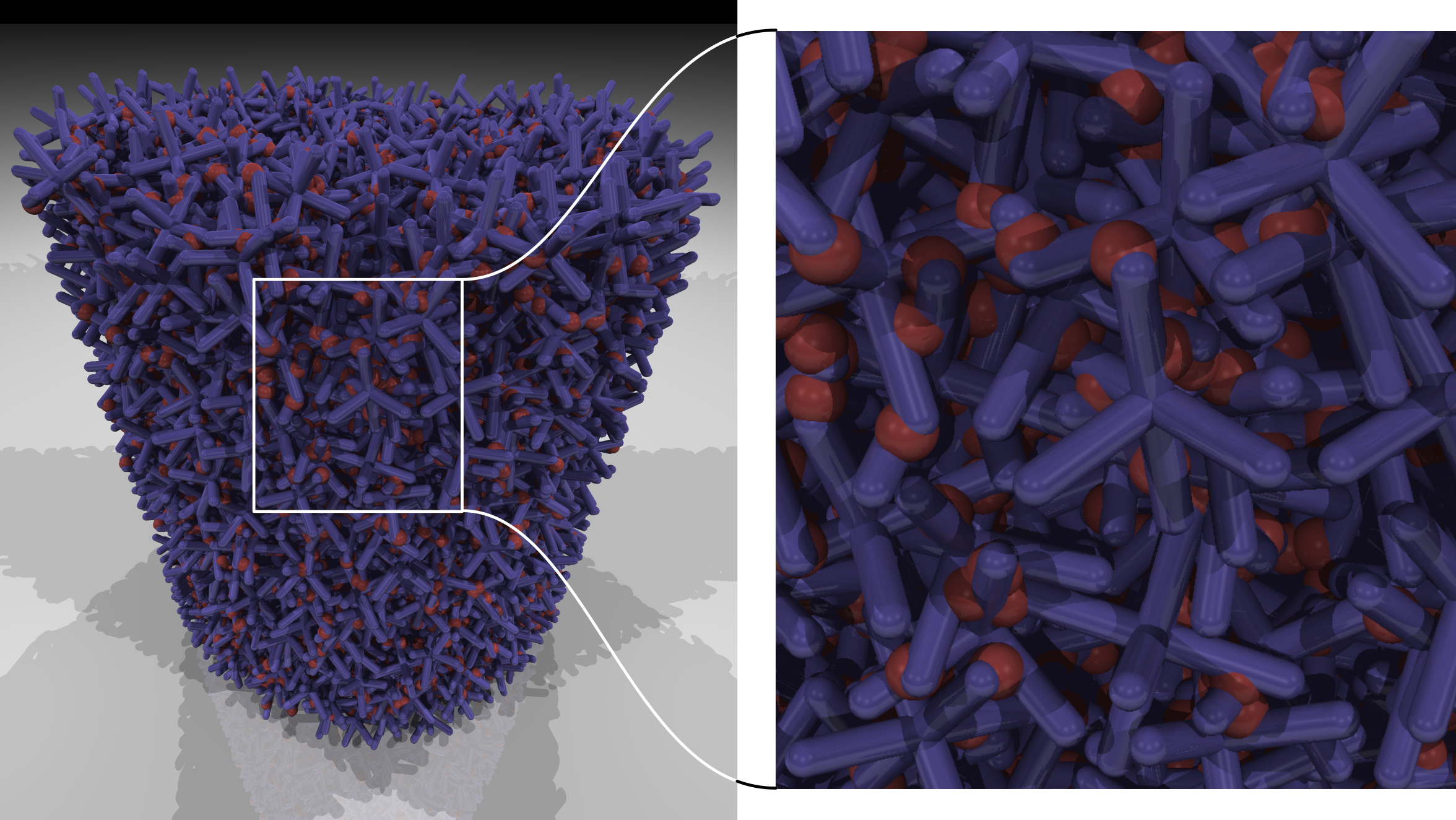}
\caption{Pile reconstruction. Left: 3D reconstructed view of a cylindrical pile composed of EPDM hexapods with a branch cross-section of $1.5$ mm. Red spheres indicate interparticle contacts. Right: Close-up view of the central region of the pile.}
\label{fig_reconst}
\end{figure}

Finally, once the positions of the particle center and the end of each branch are known, we compute, for every particle and each of its branches, the distance to all branches of neighboring particles. This is done by calculating the length of the shortest segment connecting the two line segments defined by the respective branches: the one of the particle of interest and the one of one of its neighbor. If this distance is sufficiently small, we extract a sub-cylindrical 3D matrix, $\mathcal{M}_c$, from the original full 3D image. This sub-matrix is centered on the midpoint of the joining segment and aligned along its axis. If the branches are in contact, this is precisely where the contact should occur, and large density values are expected to percolate from top to bottom within $\mathcal{M}_c$. To verify this, we analyze $\mathcal{M}_c$ column by column: for each vertical column of voxels, we record the minimum density value. This yields a 2D matrix with a disc-like structure. If any element of this matrix exceeds a density threshold $d_0$, we deduce that the particle branches are indeed in contact and can identify both the location and the specific branches involved. The scanning parameters were kept identical across all samples and were tuned to exploit the full range of the 8-bit depth ($0$ to $255$) measurement. Thus, a theoretically ideal threshold would be $d_0 \approx 127$. To allow for slight variations, we chose $d_0 = 80$ and fixed this value for all experiments. A parametric study confirmed the robustness of this choice. In Fig.~\ref{fig_reconst}, we present a constructed view of a pile made of $1.5$ mm thick hexapod of EPDM, with their contacts.

We note that in the special case of spheres ($\eta = 0$), the algorithm differs slightly, although the numerical procedures remain the same. In the end, for each prepared cylindrical pile, the experimental set-up and associated image analysis permit to obtain: 
\begin{itemize}
    \item the position of each particle,
    \item their number of branches and associated orientation, and
    \item the position of the contacts along each of these branches and the corresponding particle in contact.
\end{itemize}
These particle-scale quantities allow us, for each experiment, to construct both macroscopic and microscopic observables. These are presented in the following sections to better identify the key ingredients responsible for the emergence of geometrical cohesion and, ultimately, the stability of the particle piles.

\section{Results} \label{sec_result}

% Packing fraction
\ea{\subsection{Space filling properties}}
When studying the mechanical stability of granular systems, a key quantity of interest -- particularly due to its association with the jamming transition \cite{majmudar2007_prl,liu2010_arcmp} -- is the packing fraction. In our case, because the position of each branch of every particle is known, we compute the local packing fraction, $\phi_l$, based on the effective volume occupied by each particle and obtained via Voronoi tessellation. For concave granular systems, the construction of the tessellation is slightly modified to account for particle interlocking, as detailed in \cite{wang2024_prr}. The local packing fraction $\phi_l$ for each particle is then calculated by dividing its volume by the corresponding Voronoi cell volume.

\begin{figure}[b!]
\centering
\includegraphics[width=0.95\columnwidth]{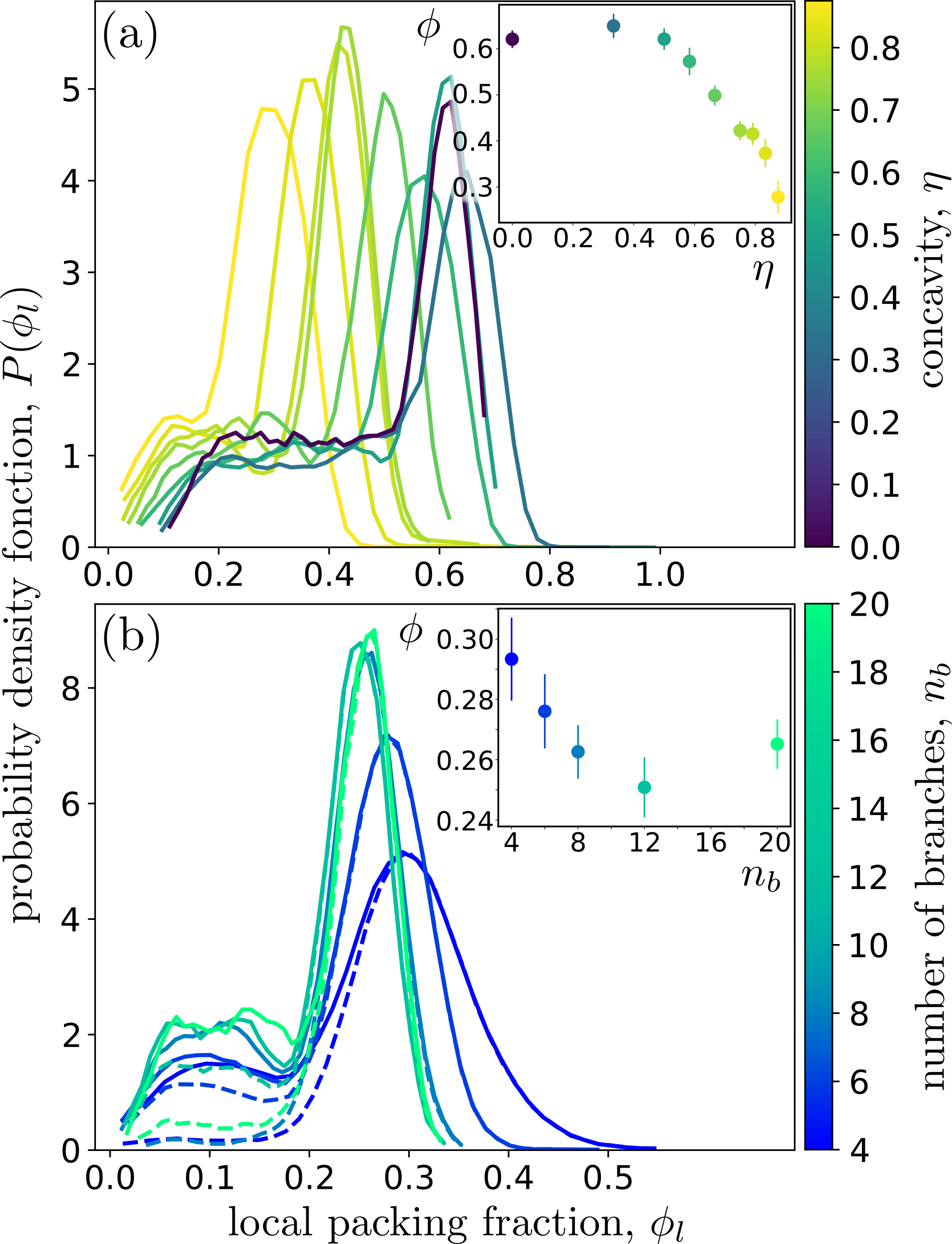}
\caption{Probability density functions (PDFs) of the local packing fraction, $\phi_l$, around a given particle are presented. Panel (a) shows the PDFs for particles with $n_b=6$ branches and varying concavities, $\eta$, while panel (b) illustrates the PDFs for particles with a concavity $\eta=0.87$ and a variable number of branches, $n_b$. In panel (b), solid lines represent the PDFs considering all particles, whereas dashed lines correspond to PDFs excluding particles near the pile boundaries. The insets display the evolution of the packing fraction, $\phi$ (calculated as average of $\phi_l$), as a function of particle concavity, $\eta$ (a), and the number of branches, $n_b$ (b). In the insets, $95~\%$ confidence intervals are shown where discernible.}
\label{fig_packing}
\end{figure}

In Fig.~\ref{fig_packing}, we present the probability density function (PDF) of the local packing fraction, $P(\phi_l)$, for columns composed of particles with varying concavity $\eta$ ($n_b=6$, confined columns because not stable) and number of branches $n_b$ ($\eta = 0.87$, unconfined columns because stable). For each configuration, the distribution typically exhibits a low-value plateau followed by a Gaussian-like peak. When $\phi_l$ values corresponding to particles near the boundaries are excluded, only the Gaussian peak remains. The position of this peak, representing the global packing of the bulk region, is tracked as a function of concavity $\eta$ in the inset of Fig.~\ref{fig_packing}a, where the global packing fraction $\phi = \langle \phi_l \rangle_{\mathrm{inner}}$ is plotted against $\eta$. For spherical particles ($\eta = 0$), $\phi$ is approximately $0.64$, close to the random close packing value \cite{torquato2000_prl}. As $\eta$ increases, $\phi$ initially rises, reaches a maximum around $\eta \approx 0.3$, and then monotonically decreases to values below $0.3$ for highly concave particles.
A similar trend is shown in the inset of Fig.~\ref{fig_packing}b, where $\phi$ is plotted as a function of the number of branches $n_b$. Here, the packing fraction initially decreases with $n_b$, reaches a minimum, and then increases again for larger $n_b$ values. \ea{Similar non-monotonic trends have been reported for a variety of non-spherical grains, reflecting a balance between improved local packing as the shape departs from sphericity and increasingly restrictive steric constraints as the geometry becomes more complex \cite{Donev2004,Sacanna2007,Cegeo2012,Azema2013b}}.

Remarkably, $\phi$ can reach values as low as $0.25$ -- indicating that approximately $75~\%$ of the pile volume is void -- and yet the pile remains mechanically stable even when unconfined. Previous works \cite{zhao2016_gm,aponte2025_gm} as well as what is shown inTab.~\ref{tab_list} show that stability, given by the collapse ratio, $\mathcal{R}$, increases monotonically with particle concavity $\eta$. However, the packing fraction $\phi$ does not vary perfectly monotonically with this parameters. As a result, there is no a clear bijective (one-to-one) relationship between stability and packing fraction. Moreover, although counterintuitive, a lower packing fraction appears to be a necessary -- but not sufficient -- condition for achieving higher stability.

\ea{\subsection{Radial distribution function and local structural organization}}
% g(r)
Packing stability is often associated with local or long-range particle ordering. For example, crystalline materials, which exhibit long-range order, are typically much stronger than amorphous or glassy systems. To better understand the origin of stability in our cylindrical piles, we investigate the extent of local ordering across different configurations. From the particle center positions, we compute and plot the average radial distribution function, $g(r/l_b)$, as a function of the normalized radial distance $r/l_b$, where $l_b = d/2 = 6$ mm is the length of a branch. This function characterizes the average local number density of particles at distance $r/l_b$, normalized by the bulk particle density \cite{torquato2002_bk}.

\begin{figure}[b!]
    \centering
    \includegraphics[width=0.95\columnwidth]{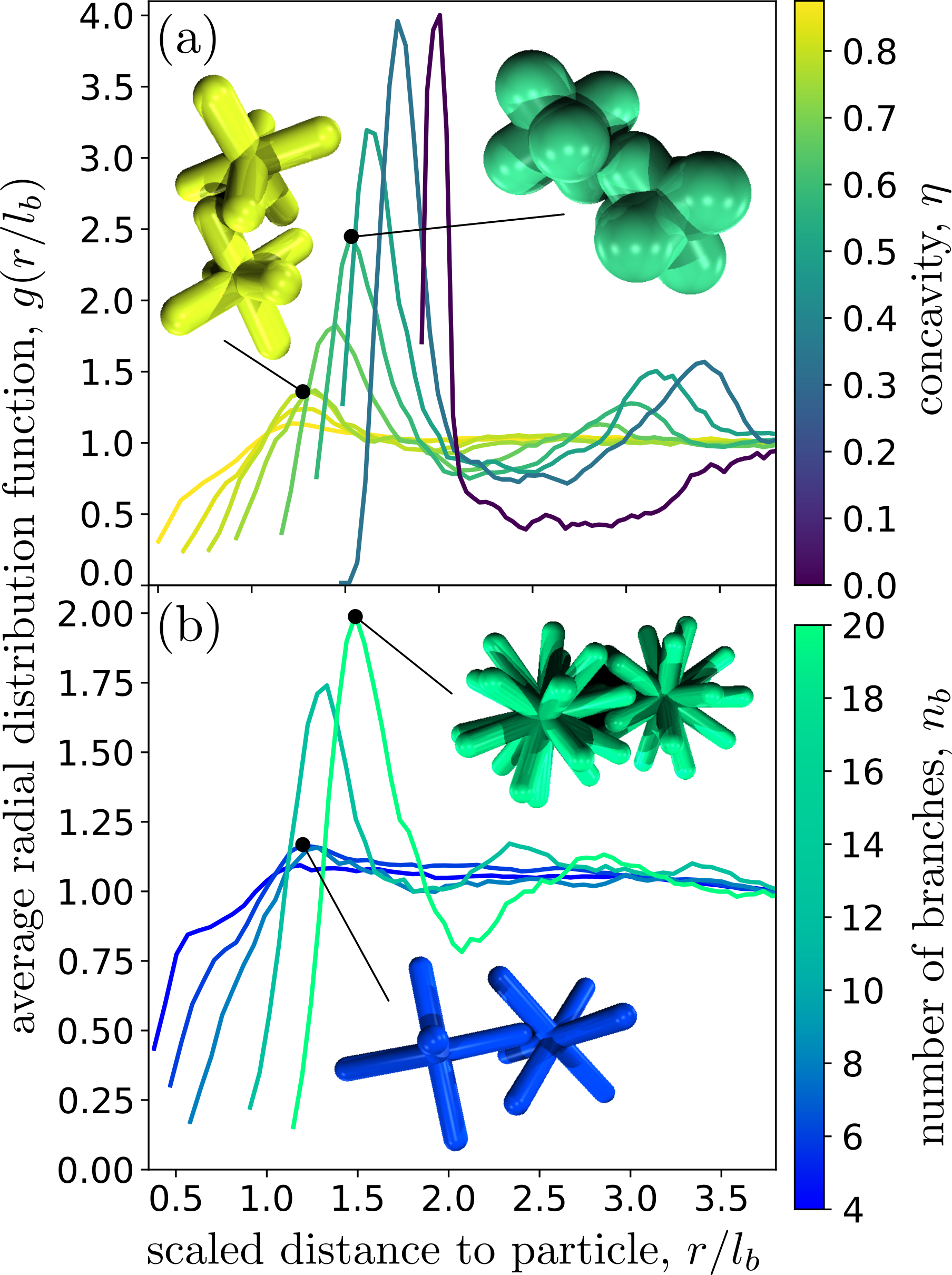}
    \caption{Evolution of the average radial distribution, $g$, as a function of the scaled distance to the particle center, $r/l_b$, is presented. Panel (a) illustrates $g(r/l_b)$ for piles composed of particles with $n_b=6$ branches and varying concavities, $\eta$, while panel (b) displays it for particles with a concavity $\eta=0.87$ and a variable number of branches, $n_b$. For $\eta = 0.79$ and $0.58$ (panel (a)) and $n_b = 6$ and $20$ (panel (b)) a pair of randomly chosen particles corresponding with the configuration at the pick are given}
    \label{fig_gr}
\end{figure}

Fig.~\ref{fig_gr}a shows $g(r/l_b)$ for varying concavity, $\eta$, and Fig.~\ref{fig_gr}b shows the same quantity as a function of the number of branches, $n_b$. In most cases, a distinct first peak is observed, corresponding to the nearest neighbors around each particle. For spheres (EPDM, $\eta = 0$), this peak is sharp and slightly above $2$, reflecting the absence of interlocking and a more ordered structure. As $\eta$ increases, the peak becomes broader and shifts to smaller $r/l_b$ values, indicating a more diffuse position of the first neighbors and particle interlocking. At low $\eta$, a second peak is also visible, corresponding to a well-organized second shell of neighbors and long range particle organization. This second peak progressively diminishes and eventually disappears as $\eta$ increases. For the highest values of $\eta$, even the first peak becomes indistinct, suggesting a near-complete loss of spatial ordering.

Based on these observations and stability results (see Tab.~\ref{tab_list} and \cite{zhao2016_gm,aponte2025_gm}), we are led to the counter-intuitive conclusion that reduced structural order is associated with increased mechanical stability -- opposite to what is typically observed in crystalline systems. However, Fig.~\ref{fig_gr}b shows that for particles with a large number of branches, the first peak is strong, and even a notable second peak appears, despite these being among the most mechanically stable configurations. This indicates that positional local order, at least as measured by $g(r/l_b)$, is not a reliable criterion for predicting the stability of packings composed of polypod particles. We note here that our measurement are in qualitative agreement with numerical simulations carried out by Conzelmann \textit{et al.} \cite{conzelmann2020_pre}.

\ea{\subsection{Orientational correlations}}
% orientation correlation
Since positional order alone does not explain stability in our system, and orientation plays a key role for anisotropic particles, we also examine orientational correlations. Indeed, due to their anisotropic shape, particle orientation also plays a key role. To assess whether local orientational ordering emerges and whether it relates to mechanical stability, we introduce a new correlation observable: $\mathcal{F}_l$, the \textit{particle orientation correlation}. This quantity measures the degree of alignment between the orientations of two given particles. This observable is constructed as follows: for two particles $i$ and $j$ having the same number of branches $n_b$, we identify the intersections of their branches with a unit sphere centered on each particle. This yields two sets of $n_b$ points, $\{P_i\}$ and $\{P_j\}$, on this unit sphere. For each point in $P_i$, we find the closest point in $P_j$ and compute the corresponding Euclidean distances. This results in a set of pairwise distances $\ell_{ij}$. The orientation correlation is then defined as the inverse of the average of these distances: $\mathcal{F}_l = 1/\langle \ell_{ij} \rangle_{ij}$. In analogy with the radial distribution function, we define $\mathcal{F}(r/l_b)$ as the average orientational correlation between particles separated by a distance $r/l_b$. To allow for comparison across different values of $n_b$, the orientation correlation $\mathcal{F}(r/l_b)$ is normalized by its long-range asymptotic value, denoted $\mathcal{F}_\infty$.

\begin{figure}[b!]
    \centering
    \includegraphics[width=0.95\columnwidth]{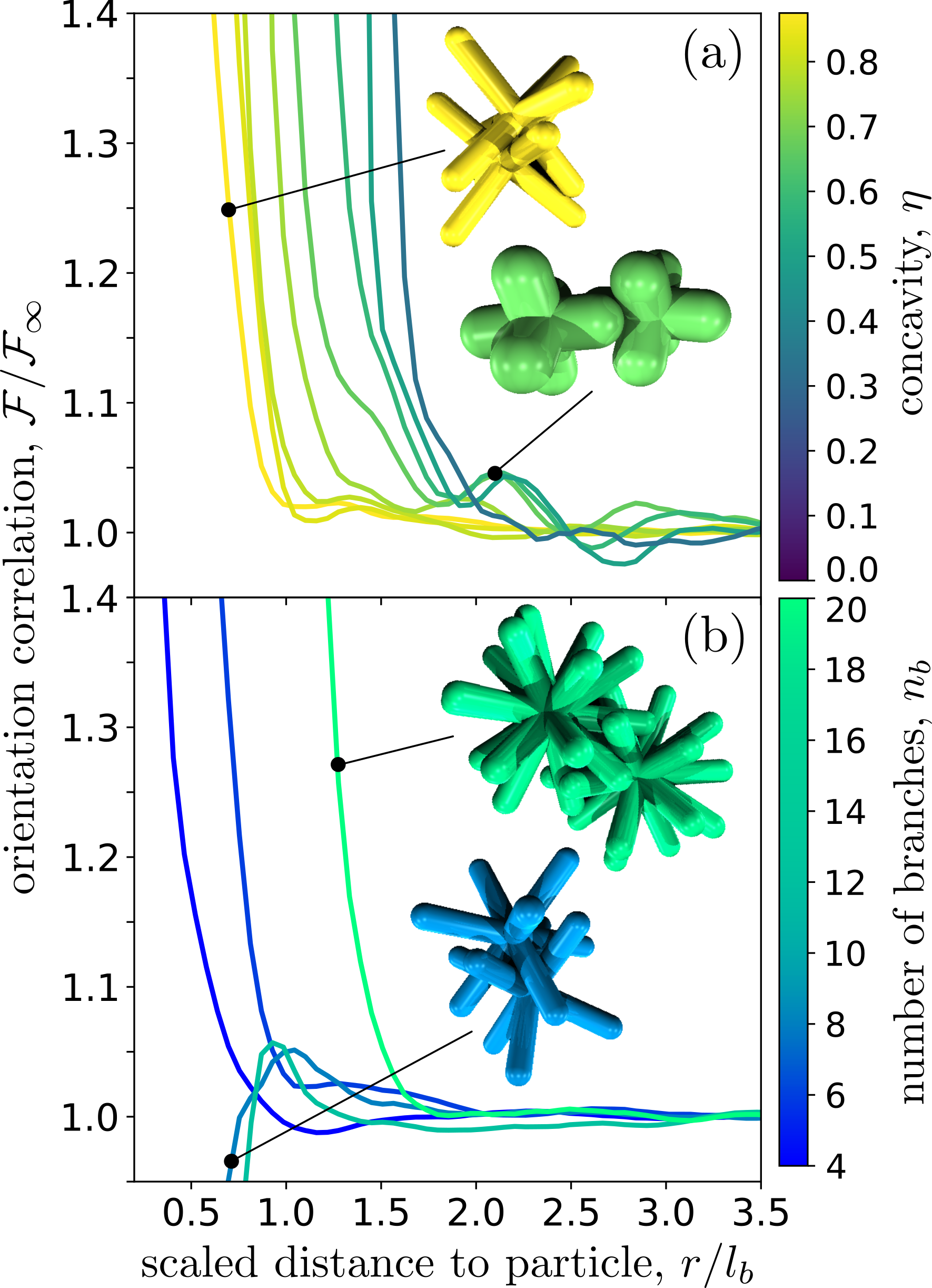}
    \caption{Evolution of the particle orientation correlation, $\mathcal{F}$, as a function of the scaled distance to the particle center, $r/l_b$, is presented. To allow for comparison across different values of $n_b$, the orientation correlation $\mathcal{F}(r/l_b)$ is normalized by its long-range asymptotic value, denoted $\mathcal{F}_\infty$. Panel (a) illustrates $\mathcal{F}(r/l_b)/\mathcal{F}_\infty$ for piles composed of particles with $n_b=6$ branches and varying concavities, $\eta$ (sphere is not presented), while panel (b) displays it for particles with a concavity $\eta=0.87$ and a variable number of branches, $n_b$. We note that this quantity cannot be computed for spheres so results are not given for $\eta=0$. For $\eta = 0.87$ and $0.67$ (panel (a)) and $n_b = 8$ and $20$ (panel (b)) a pair of randomly chosen particles corresponding with the configuration at the pick are given.}
    \label{fig_orient}
\end{figure}

Fig.~\ref{fig_orient}a shows the normalized orientation correlation, $\mathcal{F}(r/l_b)/\mathcal{F}_\infty$, for cylindrical piles composed of particles with varying concavity $\eta$, while Fig.~\ref{fig_orient}b presents the same quantity for particles with different numbers of branches $n_b$. We first observe that, depending on $n_b$, nearby particles tend to either align or anti-align their orientations. For instance, in the case of $n_b = 6$, the correlation diverges as the interparticle distance decreases, indicating a strong local alignment. In contrast, for $n_b = 8$ or $n_b = 12$, the correlation drops to zero at short distances, suggesting local anti-alignment.

Moreover, similar to the behavior observed in the radial distribution function, some orientation correlation curves exhibit one or two oscillations before stabilizing near a plateau, while others decay rapidly to their long-range asymptotic value. For piles composed of EPDM particles ($n_b = 6$), this rapid decay is observed in both limiting cases: when the particles are nearly spherical ($\eta \gtrapprox 0$) or highly concave ($\eta \approx 0.87$). This behavior is also common in most PA12-based columns, indicating the absence of significant short-range orientational order -- meaning that these systems are highly disordered. Since such orientational disorder is observed in both stable and unstable configurations, we conclude once again that mechanical stability does not originate from medium- or long-range local ordering, whether in terms of position or orientation. If anything, the results suggest quite the opposite contrary to a crystal.  

\ea{\subsection{Local connectivity:  coordination $vs$ connectivity}}
% Neighboring
Stability in granular columns is not purely a quasistatic property, it also depends on the system's ability to withstand a series of small perturbations, such as those induced when the confining tube is removed. For a column to remain stable, it must accommodate minor rearrangements without collapsing. This suggests that a key aspect of stability may lie in the capacity of a particle to rely on its close neighbors to re-establish a mechanically robust local configuration. To quantify this effect, we introduce a new local observable: the \textit{local number of neighbors}, denoted $\mathcal{N}_l$. This is defined, for a given particle, as the number of neighboring particles located within a radius of $3l_b$. To some extent, this threshold distance can be varied, but the overall trends of the results remain unchanged.

\begin{figure}[b!]
    \centering
    \includegraphics[width=0.95\columnwidth]{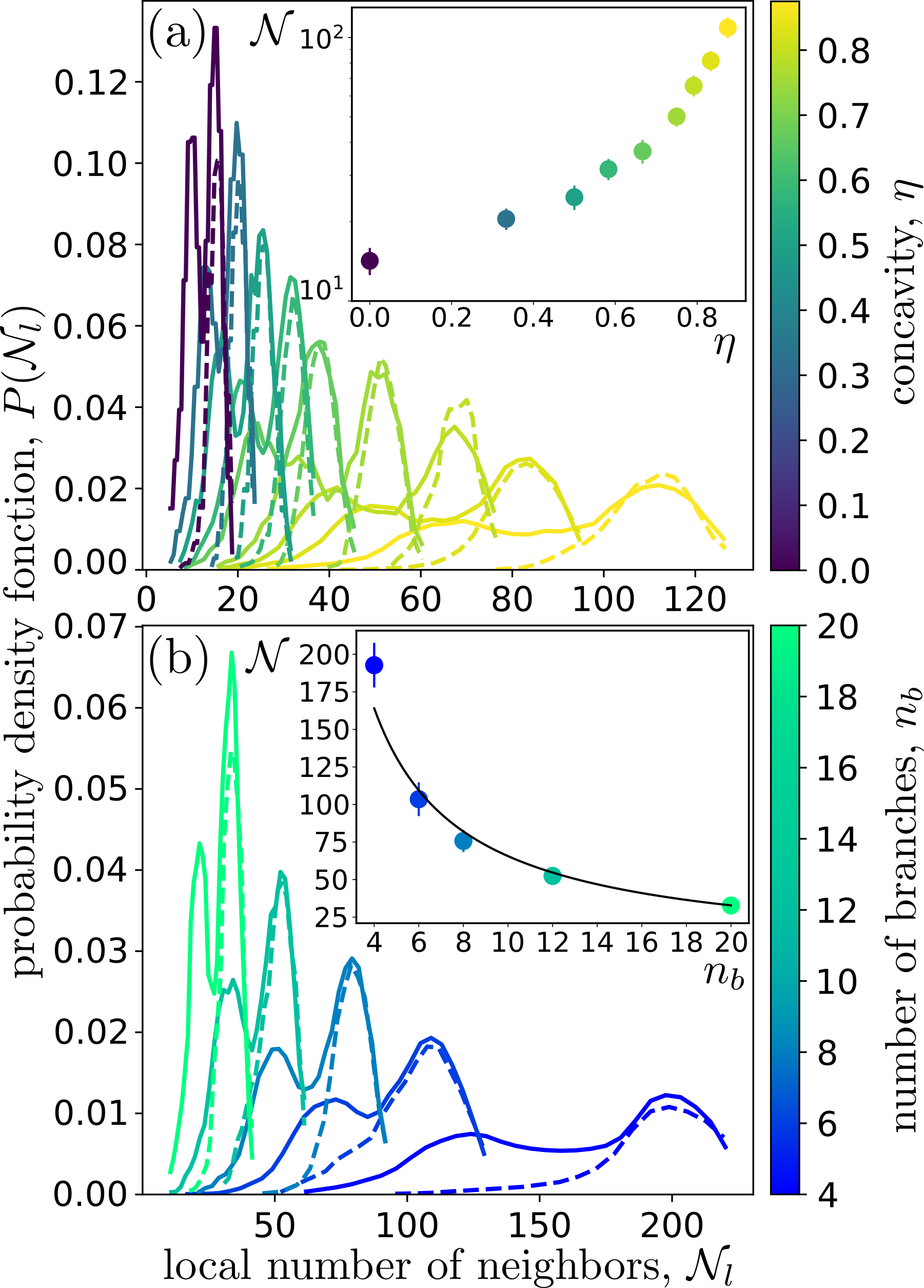}
    \caption{Probability density functions (PDFs) of the number of neighbors, $\mathcal{N}_l$, within a radius of $3 l_b$ around a given particle are presented. Panel (a) shows the PDFs for particles with $n_b=6$ branches and varying concavities, $\eta$, while panel (b) depicts the PDFs for particles with a concavity $\eta=0.87$ and a variable number of branches, $n_b$. In both cases, solid lines correspond to the PDFs considering all particles, whereas dashed lines represent PDFs excluding particles near the pile boundaries. The insets display the evolution of the average number of neighbors (away from boundaries), $\mathcal{N}$, as a function of particle concavity, $\eta$ (a), and the number of branches, $n_b$, for $\eta=0.87$ (b). In the latter case, $95~\%$ confidence intervals are shown where visible. In the inset of panel (b), the black solid line represents the inverse relationship: $\mathcal{N} = 656/n_b$.}
    \label{fig_neighbor}
\end{figure}

Fig.~\ref{fig_neighbor}a presents the PDF of $\mathcal{N}_l$ for piles composed of particles with varying concavity $\eta$, while Fig.~\ref{fig_neighbor}b shows the same observable for particles with different numbers of branches $n_b$. In each case, the PDF exhibits two peaks. The first corresponds to particles located at the pile boundaries and can be excluded from further analysis. After removing these surrounding particles, the resulting distribution becomes approximately Gaussian, from which we extract the mean and standard deviation. As shown in the inset of Fig.~\ref{fig_neighbor}a, plotting the average number of neighbors, $\mathcal{N} = \langle \mathcal{N}_l \rangle_{\mathrm{inner}}$, as a function of particle concavity $\eta$, we find that $\mathcal{N}$ increases more rapidly than an exponential function -- reaching values higher than $120$ neighbors. This rapid growth is especially striking given that it outpaces the decrease in both particle volume and global packing fraction. Additionally, the width of the distribution increases with $\eta$, indicating a growing diversity in local environments. A similar trend is observed in Fig.~\ref{fig_neighbor}b, where $\mathcal{N}$ decreases as the number of branches $n_b$ increases.

However, this variation is much weaker, approximately following an inverse relationship. Overall, we find that changes in particle proximity -- whether in trend or magnitude -- do not correlate with the mechanical stability of the packing. High or low $\mathcal{N}$ values alone do not serve as reliable indicators of structural robustness.

\begin{figure}[b!]
    \centering
    \includegraphics[width=0.95\columnwidth]{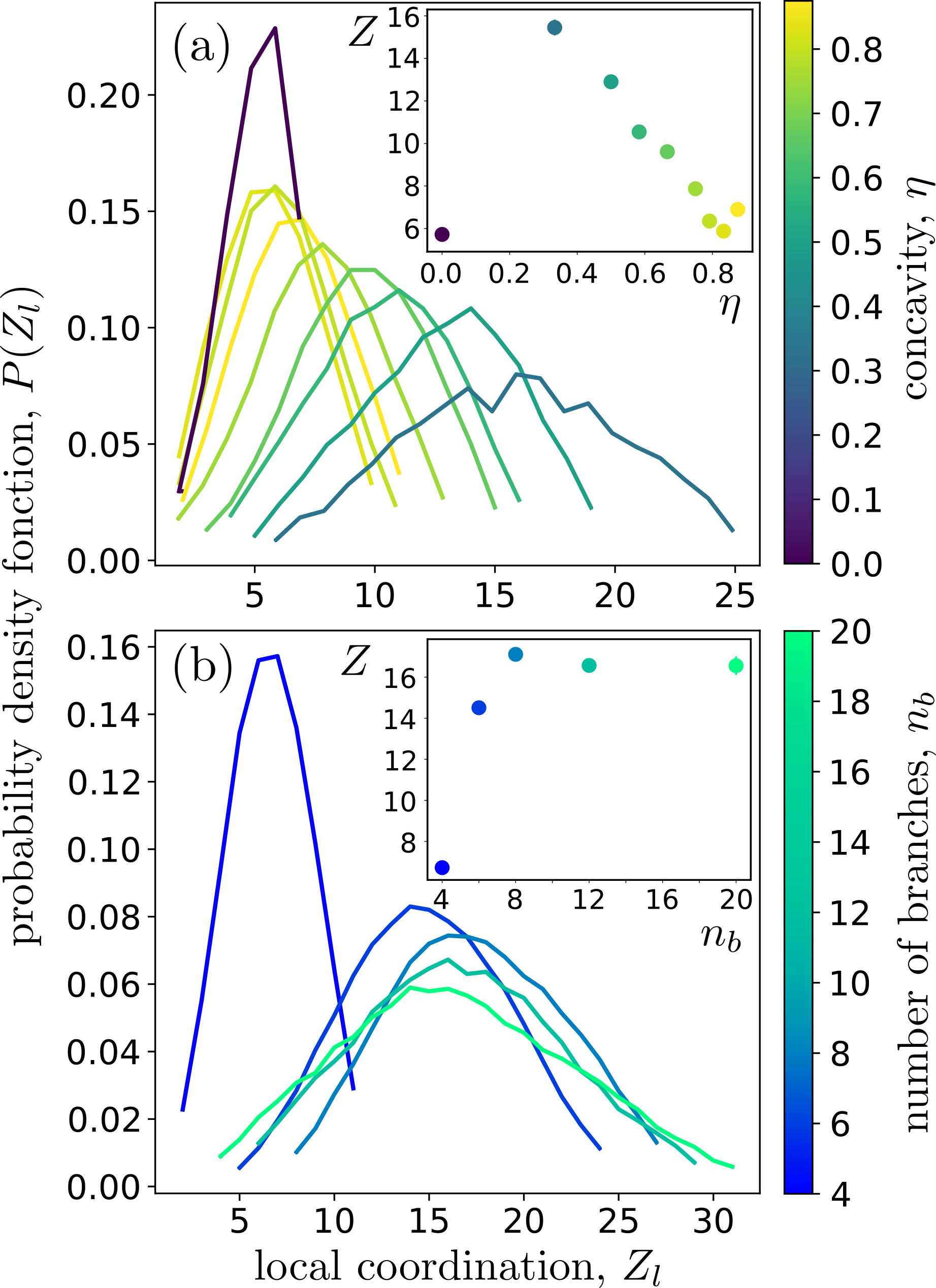}
    \caption{Probability density functions (PDFs) of the number of particle contacts, $Z_l$, are presented. Panel (a) shows the PDFs for particles with $n_b=6$ branches and varying concavities, $\eta$, while panel (b) illustrates the PDFs for particles with a concavity $\eta=0.87$ and a variable number of branches, $n_b$. The insets display the evolution of the packing coordination, $Z$ (calculated as average of $Z_l$), as a function of particle concavity, $\eta$ (a), and the number of branches, $n_b$ (b). In the insets, $95~\%$ confidence intervals are shown where discernible.}
    \label{fig_coord}
\end{figure}

% Coordination
A classical criterion for probing the stability of a granular structure is to measure its hyperstaticity. According to Maxwell \cite{maxwell1864_pm} and later Liu and Nagel \cite{liu2010_arcmp}, a three-dimensional system is jammed when the minimal number of contacts per grain exceeds $6$ for frictionless particles.
In Fig.~\ref{fig_coord}, we present the probability density function (PDF) of the number of contacts per particle, $Z_l$, for different particle types.

Fig.~\ref{fig_coord}a shows that, as expected, spheres exhibit an average coordination number $Z = \langle Z_l \rangle$ just below $6$, consistent with frictional particles. For slightly concave particles, however, $Z$ rises to very high values, up to $15$. This effect arises because adding even a small amount of concavity transforms many of the contact points that would exist on a sphere into multiple contacts involving three branches. This dramatically increases the coordination number. As $\eta$ increases further, however, $Z$ decreases approximately linearly with concavity, eventually reaching values slightly above $6$. The fact that the maximum coordination is not reached for the most concave particles is explained by the competition between two effects: although branch elongation and the number of neighbors increase with $\eta$, the packing fraction simultaneously decreases, reducing the probability of forming contacts. Importantly, the increase in average coordination is also accompanied by greater variability: the PDF of $Z_l$ broadens, and for the highest coordination cases, the packings contain both nearly isostatic particles with $6$ contacts and others with up to $25$.

Fig.~\ref{fig_coord}b shows the distribution of $Z_l$ for PA12 particles with varying numbers of branches. For the lowest case ($n_b=4$), the coordination number lies just above the isostatic threshold. As $n_b$ increases, $Z$ saturates around $16$. Again, the distribution is broad: some particles carry only $5$ contacts while others sustain up to $30$. For every cases, the lowest-contact particles are located at the system boundaries.

Taken together, these results show that coordination number alone is not a reliable stability criterion for branched particles. For instance, packings of particles with $4$ branches remain stable despite having a low coordination number ($Z=5.48$), while piles of particles with low concavity ($\eta = 0.33$) collapse even though their coordination is high ($Z=16.53$).

% Coordination and distance
The study of contact occurrence can be further refined by coupling it with the distance between contacting particles. Indeed, when a cylindrical pile is released, a contact between two particles is more likely to persist and contribute to the stability of the structure if those particles are located close to each other. In Fig.~\ref{fig_dist_cont}, for each pile, we plot the average scaled center-to-center distance between contacting particles as a function of the number of contacts shared by those particles. Results are shown for piles with different particle concavities in Fig.~\ref{fig_dist_cont}a and for particles with varying numbers of branches in Fig.~\ref{fig_dist_cont}b. In all cases, the average distance between contacting particles decreases as the number of contacts increases.

In Fig.~\ref{fig_dist_cont}a, we further observe that the contact distance systematically decreases with increasing concavity, reflecting the stronger interdigitation made possible by more concave particles. Consistently, particles sharing a larger number of contacts also display stronger interlocking. A similar trend is seen for released piles made of PA12 particles in Fig.~\ref{fig_dist_cont}b. Moreover, this latter figure reveals that the average distance between contacting particles is smaller when the number of branches is smaller, as the available space allows particles to interdigit more deeply. We deduce that a large overlapping distance between particles, by inducing more connection points, is a necessary condition to increase the likelihood of mechanical stability, but it is not sufficient on its own.

\begin{figure}[b!]
    \centering
    \includegraphics[width=0.85\columnwidth]{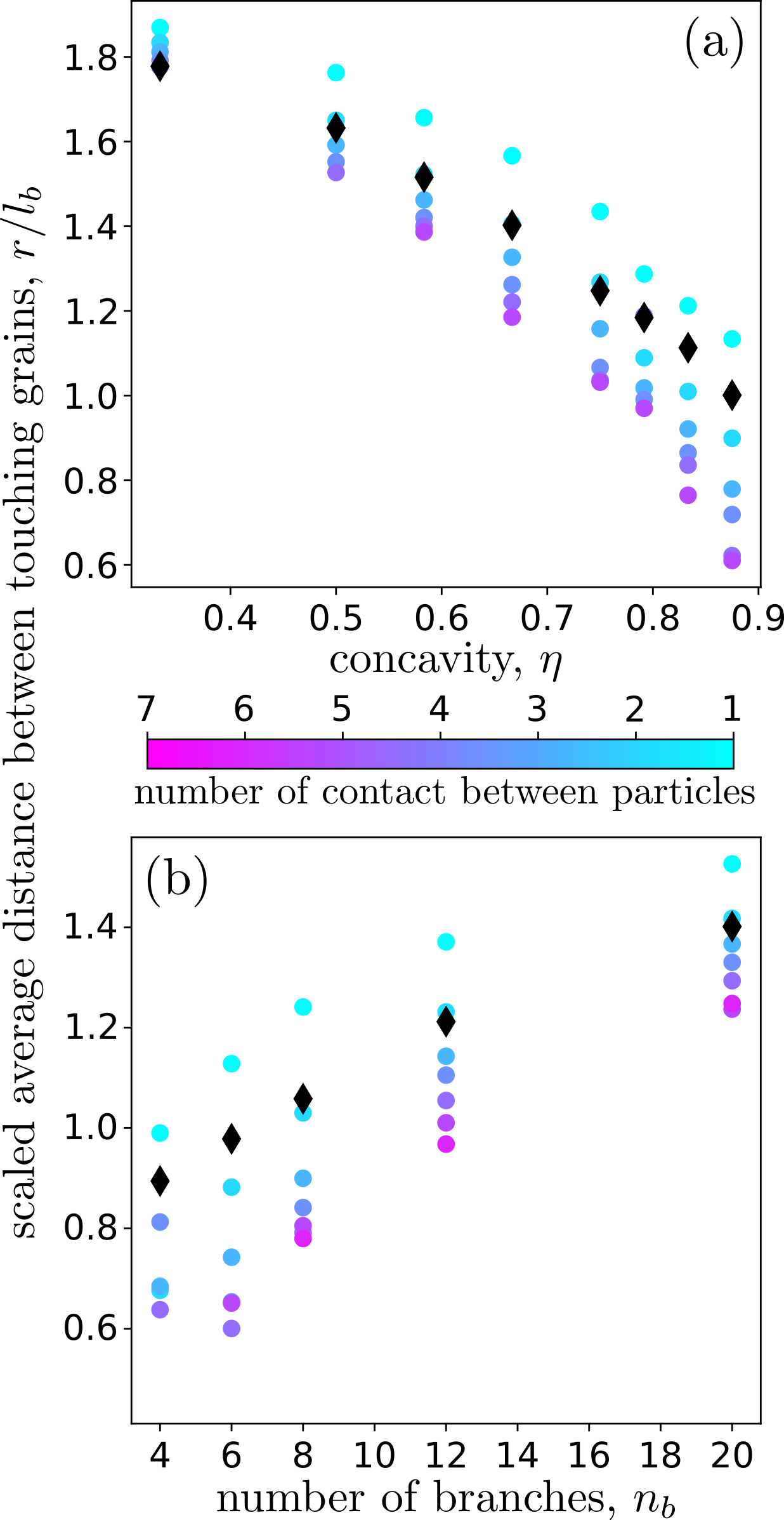}
    \caption{Average distance between contacting particles as a function of their number of contacts. Considering the number of contacts between two given particles ranging from $1$ to $7$, panel (a) presents the average scaled distance between two particles with $n_b=6$ branches as a function of concavity, $\eta$. Similarly, panel (b) displays this relationship for two particles with a concavity $\eta=0.87$ as a function of branches, $n_b$. Black diamonds indicate the average scaled distance, considering all contacting particles regardless of the number of contacts. The case of spheres is not shown, as it is trivial. Furthermore, although the scale spans from $1$ to $7$ contacts, data points corresponding to contact numbers that are topologically impossible (and therefore never observed) are omitted. 
    }
    \label{fig_dist_cont}
\end{figure}

\begin{figure}[b!]
    \centering
    \includegraphics[width=0.95\columnwidth]{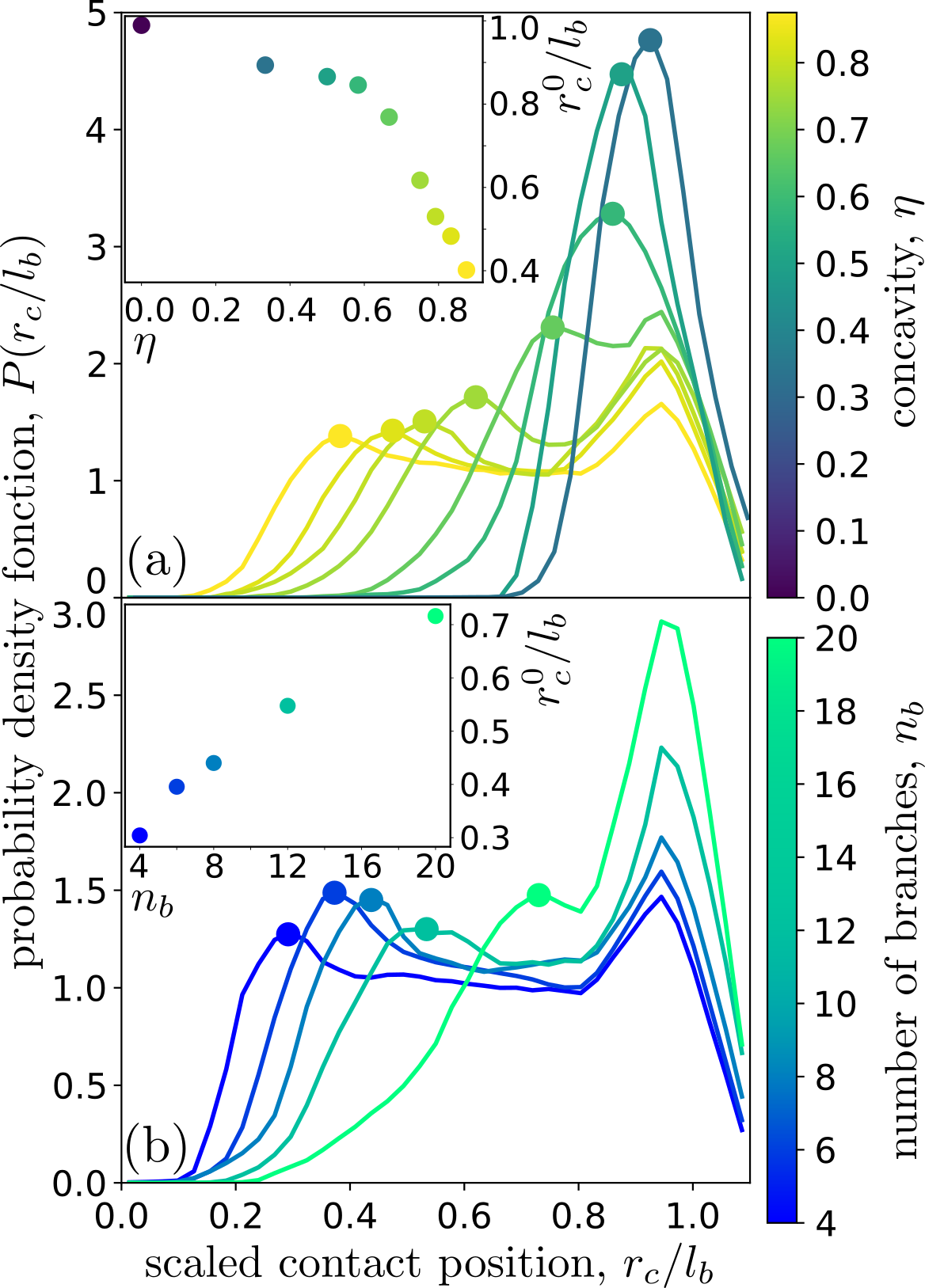}
    \caption{Probability density functions (PDFs) of the scaled contacts positions along particle branch, $r_c/l_b$, are presented. Panel (a) shows the PDFs for particles with $n_b=6$ branches and varying concavities, $\eta$, while panel (b) illustrates the PDFs for particles with a concavity $\eta=0.87$ and a variable number of branches, $n_b$. The insets display the evolution of the scaled average close contact position, $r_c^0/l_b$ (position of the first peak of $P(r_c/l_b)$), as a function of particle concavity, $\eta$ (a), and the number of branches, $n_b$ (b). In the insets, $95~\%$ confidence intervals are shown where discernible. In both main panels, the dots at the top of the peaks shows the data plotted in the inset. The case of spheres is not shown, as it is trivial.
    }
    \label{fig_pos_cont}
\end{figure}

\ea{\subsection{Contact geometry and interlocking}}
% Contact position
Then, we examine the position of contacts along the particle branches. We hypothesize that a contact is more likely to survive a small destabilization if it is closer to the particle center, particularly when it is locked between two branches. Fig.~\ref{fig_pos_cont} shows the PDF of the distance from a contact point to the particle center. Results are presented for piles composed of particles with different concavities (confined piles) in Fig.~\ref{fig_pos_cont}a and for particles with varying numbers of branches (released piles) in Fig.~\ref{fig_pos_cont}b. In most cases, two distinct peaks are observed: the farther peak corresponds to contacts at the branch extremities, which are the most likely to occur, while the closer peak corresponds to branches interlocked near the particle center of their neighbors. For particles with low concavity, the two peaks merge due to geometric proximity.

The insets of Figs.~\ref{fig_pos_cont}a and \ref{fig_pos_cont}b show the position of the closest peak as a function of concavity and number of branches, respectively. This position decreases slowly with concavity up to $\eta \approx 0.5$, after which it drops more sharply. Conversely, it increases almost linearly with the number of branches. Once again, we conclude that a significant number of contacts located close to the particle center is a necessary condition for achieving a stable packing, but it is not sufficient.

\ea{\subsection{A unified stability indicator}}
% Unified stability indicator
From the observables studied so far, and following the indications of previous work \cite{aponte2024_pre}, pile stability appears to arise from four main mechanisms:
(\textit{i}) the ability of a pile to \textit{re-stabilize} after slight perturbations, rather than its ability to simply resist them;
(\textit{ii}) strong interdigitation between particles, which maximizes the system’s capacity to reform contacts;
(\textit{iii}) significant rotational constraint between particles, which enhances contact efficiency and limits relative rotation, thereby facilitating the formation of new stabilizing contacts;
(\textit{iv}) the propensity of the system to establish frictional contacts oriented normal to the particle alignment, a configuration required for friction forces to contribute to cohesion.

Considering these four effects, we propose a new criterion to characterize the stability of a granular packing made of polypods. This criterion is constructed from the average relative branch length contained within the sphere of influence of each particle and projected along the center-center direction between particles. We denote this quantity by $\mathcal{L}$. For each particle $\mathcal{P}_0$, and for each branch of its neighbors, $\{\mathcal{P}_i\}$, we compute the portion of each branch that lies inside the sphere prescribing $\mathcal{P}_0$ (of diameter $d$). This length is normalized by $d/2$, the branch length, and projected onto the segment joining $\mathcal{P}_0$ to $\mathcal{P}_i$. The projection is explained by the fact that, since there is no hooking effect between particles, the only cohesive force between particles comes from the frictional component of the interaction forces. These latter are maximal for contact orthogonal to the $(\mathcal{P}_0,\mathcal{P}_i)$ direction. The resulting quantity is then summed over each branch of each neighbor, $\mathcal{P}_i$, and averaged over all particles, $\mathcal{P}_0$, in the packing. The value of $\mathcal{L}$ therefore quantifies the degree of interdigitation, taking into account branch alignment. To include the effect of rotational constraint, $\mathcal{L}$ is divided by the solid angle associated with a face of the Platonic solid from which the particle geometry is derived, $4\pi/n_b$. Since the only cohesive interactions in the system originate from friction, the entire quantity is finally multiplied by the interparticle friction coefficient $\mu$. We thus obtain a stability indicator $\mathcal{S}$ defined as:
\begin{equation}
    \mathcal{S} = \mu \frac{4 \pi \mathcal{L}}{n_b}
    \label{eq_crit}
\end{equation}

The stability indicator $\mathcal{S}$ is shown in Fig.~\ref{fig_dist_cont} for piles composed of particles with different concavities (confined piles) and for particles with varying numbers of branches (released piles). We first observe that $\mathcal{S}$ increases exponentially with concavity $\eta$. We also note that it increases approximately linearly with the number of branches $n_b$, up to a plateau value just below $7$. Finally, the largest $\mathcal{S}$ values measured for EPDM particles are close to the smallest values measured for PA12 particles, which is consistent with the observed differences in pile stability. As we will show in more detail later, variations of the stability indicator correlate very well with the stability of the pile.  

\begin{figure}[b!]
    \centering
    \includegraphics[width=0.75\columnwidth]{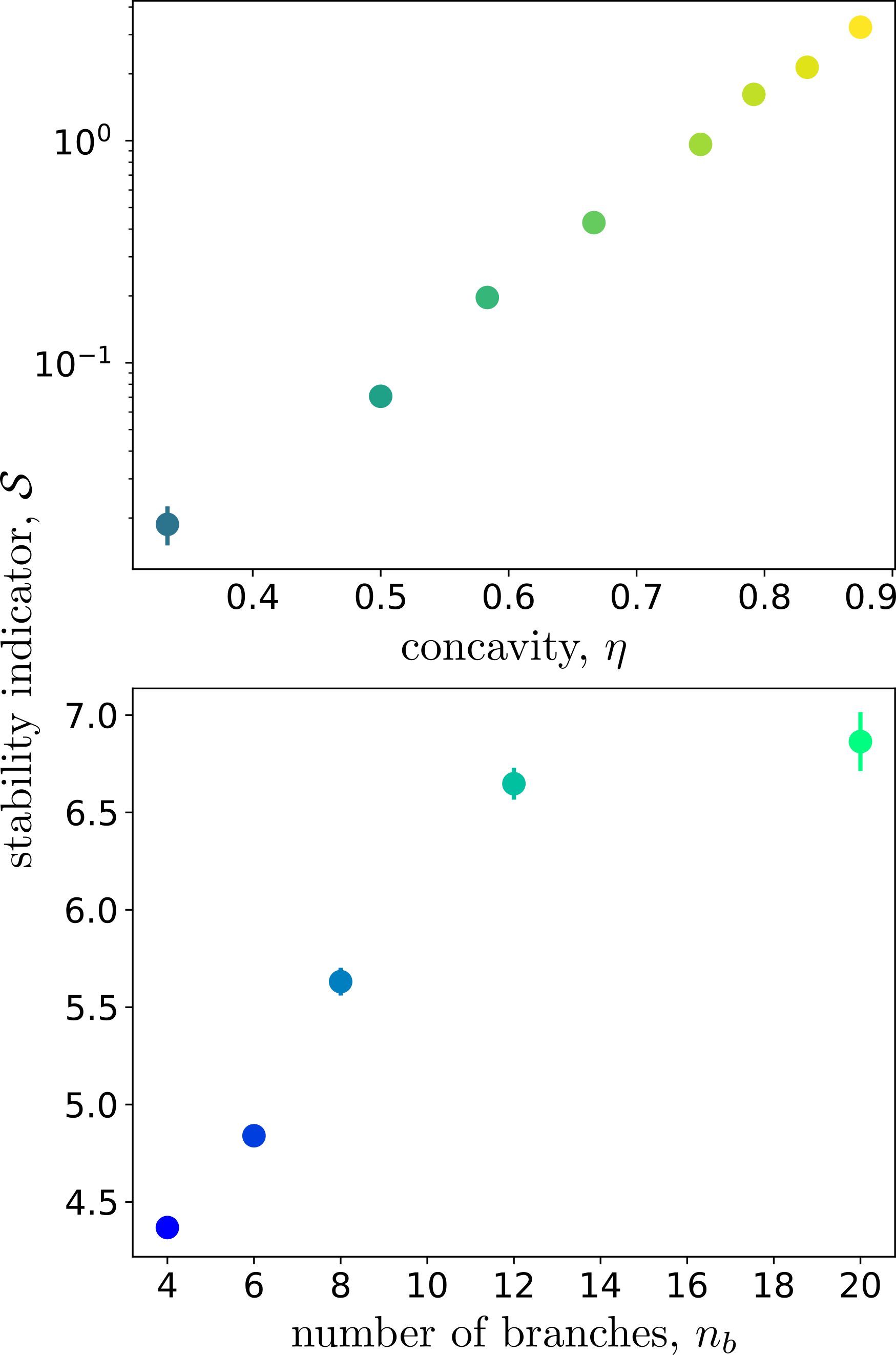}
    \caption{Stability indicator. Evolution of the stability indicator, $\mathcal{S}$ for piles made of particles with $n_b=6$ branches and varying concavities, $\eta$ (top), and for particles with a concavity $\eta=0.87$ and a variable number of branches, $n_b$ (bottom). In the top panel, the y-axis is in logarithmic scale.
    }
    \label{fig_dist_cont}
\end{figure}

% Effect of friction
Beyond the criteria used to assess pile stability, we also measured the effect of the interparticle friction coefficient on these observables for pile made of particles with the same concavity and number of branches. As shown in Tab.~\ref{tab_frict}, both the packing fraction and the number of neighbors increase significantly at low friction coefficients, but quickly saturate at higher values. Similarly, low friction facilitates particle rearrangements, allowing the system to form substantially more contacts. However, friction does not appear to affect the position of the closest contacts, which consistently remain locked near the particle centers. Still friction has a large effect of the stability indicator $\mathcal{S}$ by increasing its value linearly as expected from Eq.\ref{eq_crit}.

\begin{table}
    \begin{tabular}{l|l|l|l|l|l|}
        \cline{2-6}
                                   & $\phi$          & $Z$              & $\mathcal{N}$ & $r_c^0/l_b$ \tiny{$\times 10^{-3}$} & $\mathcal{S}$   \\ \hline
        \multicolumn{1}{|l|}{PA12} & $0.27 \pm 0.02$ & $5.48 \pm 0.08$  & $108 \pm 1$   & $393 \pm 4$                         & $5.16 \pm 0.03$ \\ \hline
        \multicolumn{1}{|l|}{EPDM} & $0.28 \pm 0.03$ & $6.89 \pm 0.08$  & $109 \pm 2$   & $398 \pm 5$                         & $3.24 \pm 0.02$ \\ \hline
        \multicolumn{1}{|l|}{HDPE} & $0.33 \pm 0.02$ & $14.39 \pm 0.11$ & $129 \pm 2$   & $368 \pm 2$                         & $1.55 \pm 0.01$ \\ \hline
    \end{tabular}
    \caption{Effect of friction. Values of the packing fraction, $\phi$, coordination, $Z$, average number of neighbors, $\mathcal{N}$, average scaled close contact position, $r_c^0/l_b$, and stability indicator, $\mathcal{S}$, for pile of particles made of PA12 ($\mu = 0.8$), EPDM ($\mu = 0.5$), and HDPE ($\mu = 0.2$) in the case of constrained not vibrated piles of particles with $6$ branches and of concavity $\eta = 0.87$.}
    \label{tab_frict}
\end{table}

% Effect of friction and vibration
Vibrating the polypod piles during preparation is also an important factor for stability. In Tab.~\ref{tab_frict}, we report the effects of vibration on the main pile observables for particles with two different friction coefficients but identical concavity and number of branches. We find that vibrations influence the packing fraction only when the friction coefficient is sufficiently high. At low friction, particles can already rearrange efficiently to reach their maximal density. A similar trend is observed for the coordination number: in highly frictional systems, vibrations markedly increase the number of contacts, whereas the increase is more modest when interparticle friction is low. For the number of neighbors, however, vibrations lead to a significant increase regardless of the friction coefficient. By contrast, we do not observe any systematic effect on the average scaled position of the closest contacts, $r_c^0/l_b$. Also vibration permits to slightly increase the stability indicator as expected from experimental collapse observation.

\begin{table}
    \begin{tabular}{ll|l|l|l|l|l|}
        \cline{3-7}
        \multicolumn{2}{l|}{}                                                             & $\phi$          & $Z$              & $\mathcal{N}$ & $r_c^0/l_b$ \tiny{$\times 10^{-3}$}  & $\mathcal{S}$   \\ \hline
        \multicolumn{1}{|l|}{\multirow{2}{*}{\rotatebox{90}{\tiny{EPDM}}}} & vibrated     & $0.26 \pm 0.04$ & $13.99 \pm 0.12$ & $116 \pm 1$   & $380 \pm 1$                          & $3.29 \pm 0.03$ \\ \cline{2-7} 
        \multicolumn{1}{|l|}{}                                             & not vib.     & $0.28 \pm 0.03$ & $6.89 \pm 0.08$  & $109 \pm 2$   & $398 \pm 2$                          & $3.24 \pm 0.02$ \\ \hline
        \multicolumn{1}{|l|}{\multirow{2}{*}{\rotatebox{90}{\tiny{HDPE}}}} & vibrated     & $0.33 \pm 0.02$ & $15.63 \pm 0.10$ & $138 \pm 2$   & $365 \pm 1$                          & $1.64 \pm 0.01$ \\ \cline{2-7} 
        \multicolumn{1}{|l|}{}                                             & not vib.     & $0.33 \pm 0.02$ & $14.39 \pm 0.11$ & $129 \pm 2$   & $368 \pm 1$                          & $1.55 \pm 0.01$ \\ \hline
    \end{tabular}
    \caption{Effect of friction and vibration. Values of the packing fraction, $\phi$, coordination, $Z$, average number of neighbors, $\mathcal{N}$, average scaled close contact position, $r_c^0/l_b$, and stability indicator, $\mathcal{S}$, for pile of particles made of EPDM ($\mu = 0.5$) and HDPE ($\mu = 0.2$) in the case of vibrated and not vibrated piles of particles with $6$ branches and of concavity $\eta = 0.87$.}
    \label{tab_vib}
\end{table}

\section{Concluding discussions} \label{sec_ccl}

The experiments presented in this study provide local, particle-scale measurements of cylindrical polypod piles, while systematically varying particle frictional properties, geometry, and packing preparation -- parameters that directly influence stability. This approach allows us to identify the key observables underlying \textit{geometrically induced stability} and to explain the origin of this mechanical property. The overarching goal is to determine, from the local microstructure of a pile, whether it is more likely to collapse or to remain standing once confining walls are removed. Such an understanding would enable the construction of a family of observables capable of characterizing the strength of structures stabilized by particle geometry. The present results represent a major step in this direction. 

\begin{figure}[b!]
    \centering
    \includegraphics[width=0.7\columnwidth]{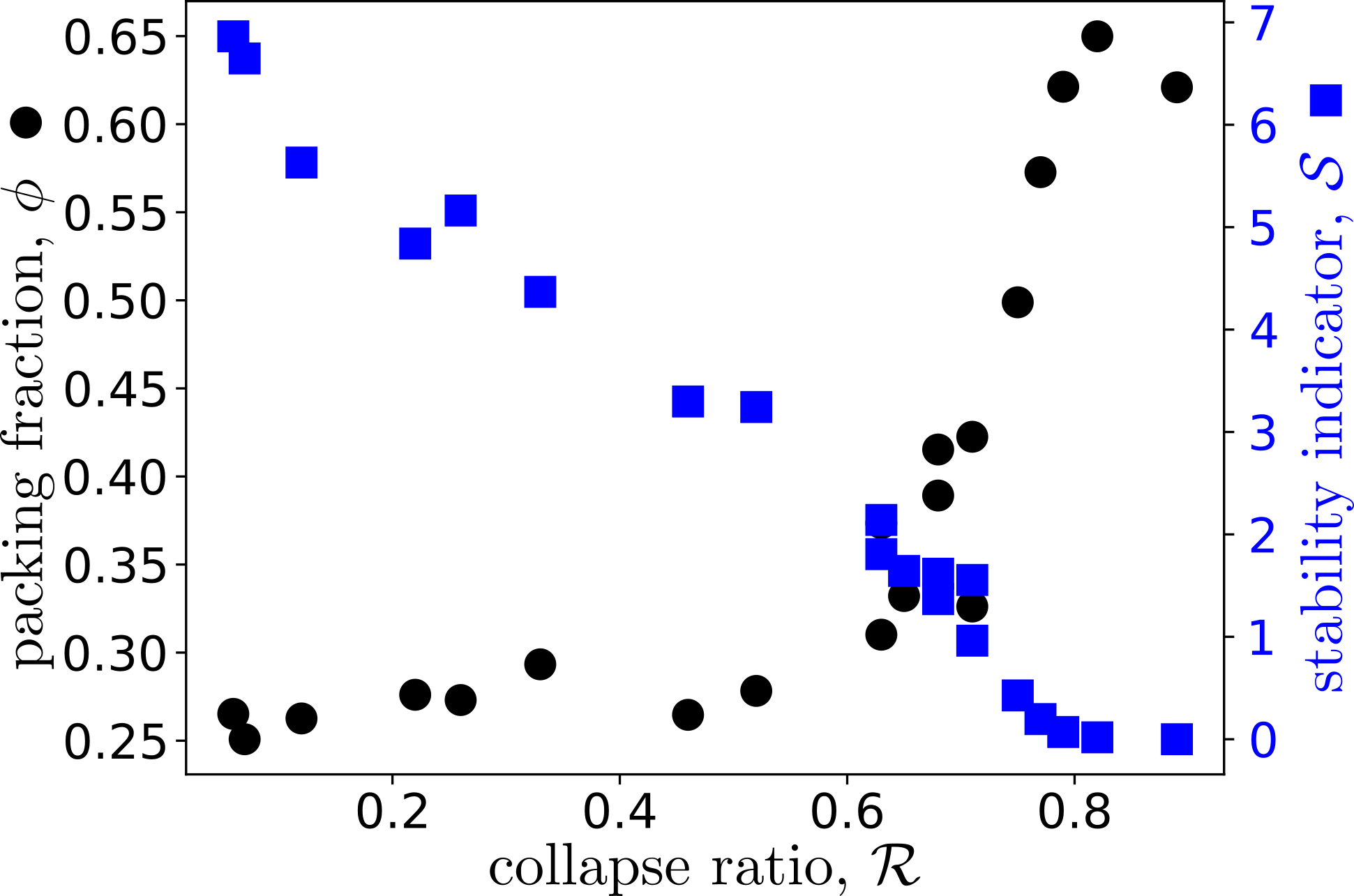}
    \caption{Correlation with collapse ratio. Evolution of the stability indicator (blue squares), $\mathcal{S}$, and of the packing fraction (black circles), $\phi$, as a function of the collapse ratio $\mathcal{R}$ for all experiments. Each point corresponds to a single experiment. Error bars are omitted for clarity.
    }
    \label{fig_check}
\end{figure}

First, we note that a low packing fraction appears to be a necessary condition for observing a stable structure. However, it is not sufficient, since, as illustrated in Fig.\ref{fig_check}, some systems with low packing fraction remain less or unstable. At first sight, the observation that lower density is associated with higher stability may seem paradoxical, as lower density is usually linked with weaker structures. This apparent paradox arises from the fact that the strength of these systems does not result from a regular, highly ordered packing -- which would indeed yield higher density -- but rather from a more disordered arrangement. In particular, long-range positional ordering, as captured by the radial distribution function $g(r)$, is not correlated with greater stability; if anything, the opposite trend is observed.

A similar conclusion holds when considering particle orientation correlations, $\mathcal{F}(r)$. Long-range orientational correlations -- or anticorrelations -- such as extended nematic order, are not associated with enhanced geometrical cohesion. Consequently, studies focusing on crystalline order in metagrains \cite{miszta2011_nm,meng2020_part,stannarius2022_gm} do not provide relevant insights into geometrically induced cohesion. Instead, cohesion in these systems emerges from structural randomness.

Coordination number, which is often considered the most reliable criterion for assessing the stability of granular packings, also fails to fully predict the strength of polypod structures. A high coordination number indeed increases the likelihood of stability, but once again it is not a sufficient condition: it reflects the number of existing contacts, yet provides no information about the system’s ability to form new contacts when the packing is destabilized. By combining information on both the number of contacts and the distances between particles, we find that the most significant interactions are those involving particles that share multiple contacts, as these particles interlock more strongly. This conclusion is reinforced by examining the average position of contacts along the branches: the most stable packings are those in which a large fraction of contacts are locked close to the particle center. Such central locking appears to be a necessary condition for stability -- yet, crucially, it remains insufficient on its own.

These observations suggest that the strength of a structure composed of non-convex particles -- its \textit{geometrical cohesion} -- emerges from a competition between two mechanisms. On the one hand, particles must be able to approach each other closely despite branch congestion: this is the role of \textit{particle interdigitation}. On the other hand, branches must align in such a way that they restrict relative rotations: this is the \textit{inter-particle rotational constraint}. If branches are too thick, particles can neither come close to one another nor effectively constrain rotation. If branches are thin enough, both mechanisms become possible, but then the number of branches becomes the decisive factor: with too few branches, rotational constraint is insufficient; with too many, particles can no longer approach closely enough to interdigitate effectively.

Moreover, because neither interlocking nor entanglement can occur with these particle families, the only forces capable of pulling particles toward each other are the tangential components of contact forces, \textit{i.e.}, the friction forces. Geometrical cohesion therefore also depends on the ability of the system to form frictional contacts oriented perpendicular to the center-to-center alignment of the particles.

Alse, as in the case of staples \cite{pezeshki2025_jmps}, we postulate that the cohesion of structures composed of non-affine particles emerges from a dynamical process. Final stability is not governed by a static contact network, but rather by the continuous breaking and rapid reformation of contact chains that collectively stabilize the system. This mechanism is reminiscent of the processes believed to underlie the long-term stability of ancient dry-stone walls \cite{colas2010_es}. In this view, determining whether a system can be stable reduces to evaluating the number and robustness of potentially reformable contact chains. Stability is therefore not a fixed state but a process of assembly, disassembly, and re-conformability -- a form of structural recyclability. This perspective also explains why coordination number alone is less meaningful for characterizing pile stability than other observables such as the number of neighbors or the average position of contacts along branches, both of which more directly quantify the capacity for local stable reconfigurations. 

Taken together -- interdigitation, rotational constraint, friction-mediated cohesion, and the ability of a pile to \textit{re-stabilize} after small perturbations rather than merely resist them -- these mechanisms explain the origin of \textit{geometrical cohesion}.

These different ingredients are captured by the stability indicator $\mathcal{S}$. As shown in Fig.~\ref{fig_check}, $\mathcal{S}$ is strongly correlated with stability as quantified by the collapse ratio $\mathcal{R}$. This not only validates $\mathcal{S}$ as a powerful tool to measure, predict, and quantify the stability of polypod packings, but -- more importantly -- it confirms that geometrical cohesion arises from:
\begin{itemize}
    \item the ability of a pile to \textit{re-stabilize} after slight perturbations rather than merely resist them;
    \item strong interdigitation between particles, which maximizes the system’s capacity to reform contacts;
    \item significant rotational constraint between particles, which enhances contact efficiency and limits relative rotation, thereby facilitating the formation of new stabilizing contacts;
    \item the propensity of the system to establish frictional contacts oriented normal to the particle alignment, a configuration required for friction to effectively contribute to cohesion.
\end{itemize}

These results provide objective guidelines for designing stable granular packings of arbitrary polypod shape capable of withstanding diverse loading conditions. For instance, to obtain a highly stable system, it is far more effective to increase branch length (or reduce branch thickness) -- which causes the stability indicator to grow exponentially -- than to increase interparticle friction or the number of branches, which cause $\mathcal{S}$ to increase only linearly at best. Moreover, since $\mathcal{S}$ can be evaluated locally, it becomes possible, for packings of more complex shapes than cylinders, to identify weak regions that may trigger the collapse of the entire structure. This would make $\mathcal{S}(\vec{r})$ a new physical observable capable of quantifying the local mechanical strength of a granular system, in much the same way that the Young’s modulus characterizes the stiffness of a bulk material. Finally, because many other families of non-interlocking concave particles exist, it would be of great interest to examine whether our stability indicator can be generalized or extended to broader classes of particle geometries.

\begin{acknowledgments}
The authors acknowledge financial support from ANR MICROGRAM (ANR-20-CE92-0009). We would also like to express our gratitude to Gille Camp and Renaud Lebrun for their technical assistance in setting up the experiment. We acknowledge the imaging facility MRI, member of the national infrastructure France-BioImaging supported by the French National Research Agency (ANR-24-INBS-0005 FBI BIOGEN). Finally, we wish to express our deepest gratitude and respect to Christian Barbéris from the CFO company (Conception Fabrication Outillage, Ribaute-les-Tavernes, France). With his great expertise, he conceived and fabricated the mold used to produce the particles for this study. Sadly, he passed away in September 2025.
\end{acknowledgments}

\bibliography{biblio}

\end{document}